\documentclass[aps,prl,twocolumn,nofootinbib,groupedaddress,amsfonts,floatfix]{revtex4} 
\usepackage{graphicx,amsmath,amssymb,amstext}
\usepackage{amssymb,amsbsy,amsfonts,amsthm,color}

\usepackage{epsfig}
\usepackage{graphicx}
\usepackage{subfigure}

\usepackage{color}
\newcommand{\Beq}{\begin{eqnarray}}
\newcommand{\Eeq}{\end{eqnarray}}

\newcommand{\Aint}{\mathcal{A}_{\rm{int}}}

\newcommand{\eqn}[1]{Eq. (\ref{#1})}
\newcommand{\nn}{\nonumber \\}

\newcommand{\killspace}{\vskip -5pt}

\begin{document}

\title{A Clash of Kinks: Phase shifts in colliding non-integrable solitons}
\date{\today}

\author{Mustafa A. Amin${}^1$}
\author{Eugene A. Lim${}^2$}
\author{I-Sheng Yang${}^3$}

\affiliation{${}^1$Kavli Institute for Cosmology and Institute of Astronomy, Madingley Rd, Cambridge CB3 0HA, United Kingdom}
\affiliation{${}^2$Theoretical Particle Physics and Cosmology Group, Physics Department,
Kings' College London, Strand, London WC2R 2LS, United Kingdom}
\affiliation{${}^3$IOP and GRAPPA, Universiteit van Amsterdam, Science Park 904, 1090 GL Amsterdam, Netherlands}

\begin{abstract}
We derive a closed-form expression for the phase shift experienced by 1+1 dimensional kinks colliding at ultra-relativistic velocities ($\gamma v\gg 1$), valid for arbitrary periodic potentials. Our closed-form expression is the leading order result of a more general scattering theory of solitary waves described in a companion paper \cite{ALYlong}. This theory relies on a small kinematic parameter $1/(\gamma v)\ll 1$ rather than a small parameter  in the Lagrangian. Our analytic results can be directly extracted from the Lagrangian without solving the equation of motion. Based on our closed-form expression, we prove that kink-kink and kink-antikink collisions have \emph{identical} phase shifts at leading order. 
\end{abstract}

\maketitle
\section{Introduction and Summary}
\killspace

Scattering theory provides a crucial link between the Lagrangian specifying the properties of fields and outcomes of experiments. The scattering theory of solitary waves (localized waves that travel without distortion -- sometimes colloquially called solitons \cite{Scott:1973eg}) is particularly interesting since such waves play an important role in many disparate fields from modeling fluxons in superconductivity \cite{fultondynesanderson73}, optics \cite{BulCau80}, high energy physics (e.g. \cite{Wei12}) to cosmology \cite{EasGib09,AguJoh09,GibLam10,Kle11,Amin:2011hj}.  While isolated solitary waves have been studied in detail, the physics of their scattering is not well understood analytically, except in special integrable cases.

In this paper, we will consider the simplest of the such solitary waves, that of a single canonical scalar field governed by a periodic potential in (1+1) dimensions,
\begin{eqnarray}
\mathcal{L} &=& \frac{1}{2}(\partial_t\phi)^2-\frac{1}{2}(\partial_x\phi)^2 -V(\phi)~, 
\label{eq-L} \\
V(\phi) &=& V(\phi+\Delta\phi)~.  \label{eq-V}
\end{eqnarray}
The equation of motion for $\phi$ is
\Beq
\partial_t^2\phi-\partial_x^2\phi+V'(\phi)=0~. \label{eq-EOM}
\Eeq
The potential $V(\phi)$ is a general periodic potential with multiple minima. A simple solitary wave in this theory is a kink: an interpolation between two adjacent minima which can travel at a constant velocity without any distortion. Apart from an isolated special case with $V(\phi)= (1-\cos \phi)$, known as the Sine-Gordon model, the usual way to predict the outcome of collisions between kinks is to numerically evolve the equation of motion (e.g. \cite{PhysRevLett.98.104103,PhysRevLett.107.091602}). On the analytical side, perturbative predictions have been only done for cases which are close the Sine-Gordon case (see reviews \cite{Mal85b,McLSco78}).

In this paper we demonstrate an analytical calculation for the outcome of collisions between kinks. Our main result is a closed form expression for the phase shift (spatial translation) experienced by a stationary kink due to a collision with an incoming kink/anti-kink with velocity $v\rightarrow 1$:
\begin{eqnarray}
\Delta x \label{eq-shift}
&=&\frac{1}{2(\gamma v) M}\int_{0}^{\Delta\phi}\int_{0}^{\Delta\phi}d\phi_1d\phi_2 \\ 
& & \left[\frac{V(\phi_1)+V(\phi_2)-V(\phi_1+\phi_2)}{\sqrt{V(\phi_1)V(\phi_2)}}\right]+\mathcal{O}[(\gamma v)^{-2}]~,
\nonumber
\end{eqnarray}
where $\gamma=(1-v^2)^{-1/2}$ and $M = \int_0^{\Delta\phi} d\phi \sqrt{2V(\phi)}$ is the energy of the stationary kink.   

Remarkably,  the phase shift is {\em an explicit function of the incoming velocity of the colliding wave and the potential, and does not require evaluation of the equation of motion.} We emphasize that the periodic potential $V(\phi)$ need not be perturbatively close to the Sine-Gordon case and hence encompasses a much wider class of models compared to previous studies \cite{Mal85b}. 

This simple form of \eqn{eq-shift}  allows us to immediately draw several insights into the nature of kink interaction.  First, {\em the leading order phase shift is the same for both kink-kink and kink-antikink collision}\footnote{While there is no contradiction with our result, it is known that at long ranges kink-kink interaction is repulsive but kink-antikink interaction is attractive \cite{Manton:1978gf,Mantonbook,Wei12}.}. Second, \eqn{eq-shift} is not positive definite hence there exists models with negative phase shifts. In particular, this implies that there exists an entire class of models where $\Delta x=0$. Third, even though the collision is dissipative, the lack of time dependence in the right hand side of \eqn{eq-shift} at leading order implies that  {\em no velocity change} occurs at this order. Hence the collision is almost elastic.

Eq.~(\ref{eq-shift}) is the leading order result of a perturbative expansion in $(\gamma v)^{-1}$ described in the companion paper \cite{ALYlong} which applies to a wider class of solitary waves, and includes a prescription for an order by order calculation of higher order terms. Here we focus on the derivation of this leading order result for the phase shift. As a check, we performed detailed numerical simulations and found excellent agreement with our analytic answer.

\section{Deriving the phase shift}
\killspace

\begin{figure}
\begin{center}
\includegraphics[width=5cm]{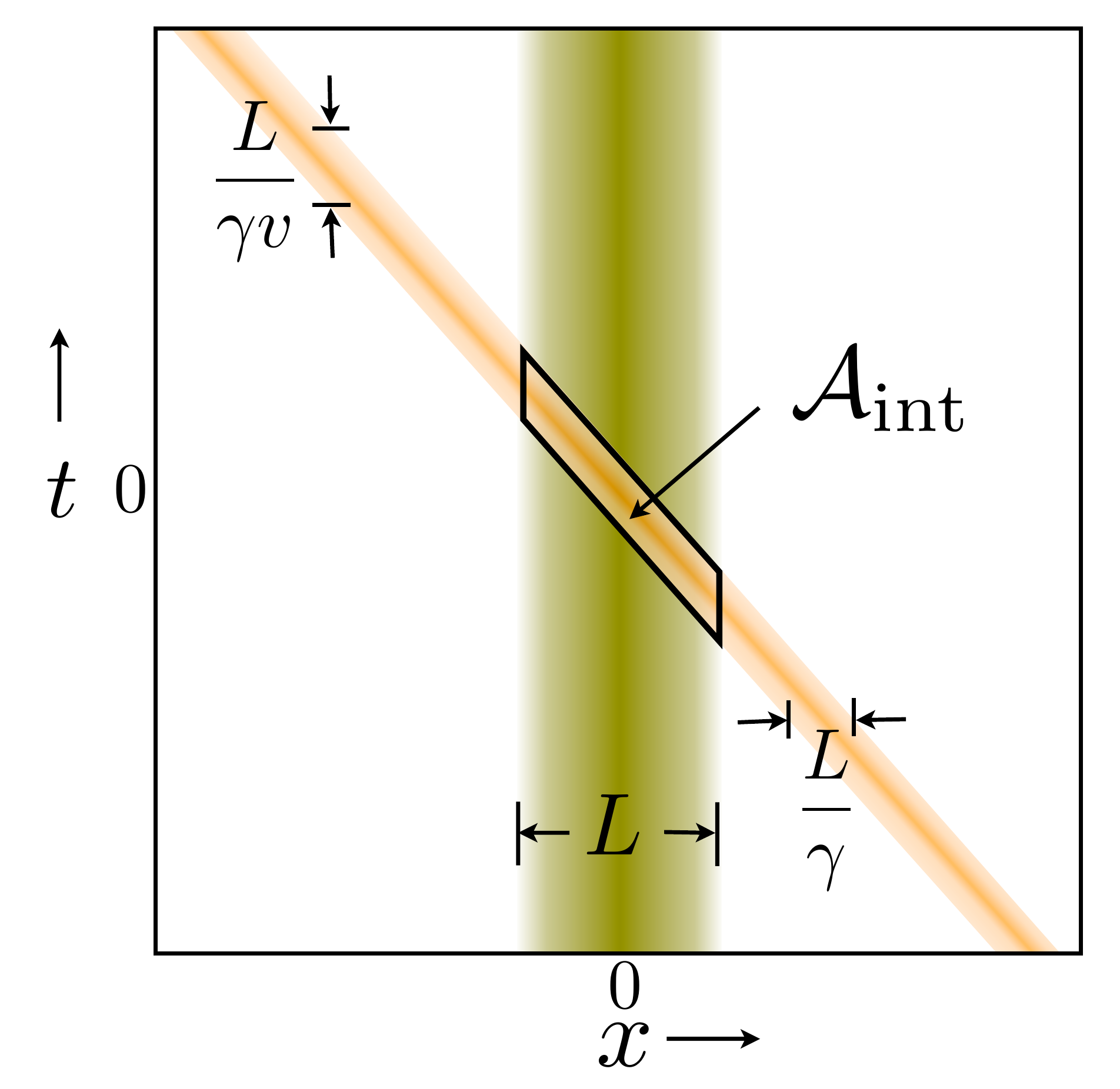}
\caption{The overall space-time are occupied by the two solitary waves, i.e. the area where their field values deviate significantly from vacuum is shown above. The green strip represents the stationary solitary wave, whereas the orange one is the incoming solitary wave. Assuming that the solitary waves simply pass through each other, the area of overlap between the solitary waves $\Aint$ is denoted by the black parallelogram. Elementary geometry yields $\Aint\propto1/(\gamma v)$. For ultra relativistic collisions, the Lorentz contraction of the incoming solitary waves as well as the short time of collision are responsible a suppressed $\Aint$.
\label{fig-IntArea}}
\end{center}
\vskip -20pt
\end{figure}

\begin{figure}
\begin{center}
\includegraphics[width=8.5cm]{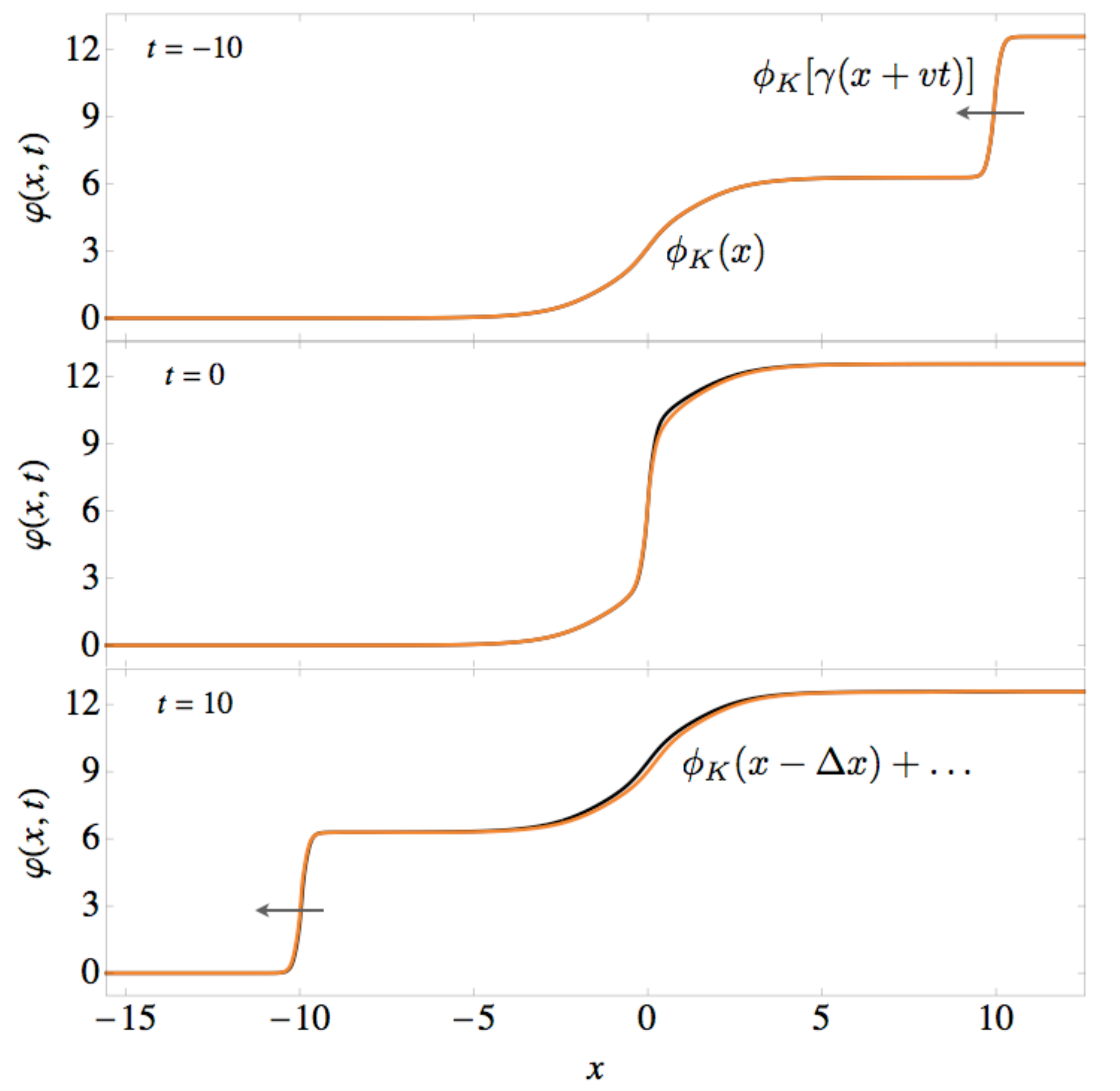}
\caption{The orange curves are the numerical field profiles before, during, and after collision. The black curve is the superposition solution which ignores all interactions. We can see a clear phase shift after the collision.
\label{fig-FieldCollision}}
\end{center}
\vskip -20pt
\end{figure}

A stationary kink $\phi_K(x)$ is a solution to the equation of motion for the Lagrangian in ~\eqn{eq-L},
\begin{equation}
\phi_K''(x) = V'[\phi_K(x)]~,
\label{eq-kink}
\end{equation}
such that $\phi_K(-\infty)=0$ and $\phi_K(\infty)=\Delta\phi$. The profile exponentially approaches vacuum values beyond some region $\sim L$ from its center -- see the part of the curve labeled $\phi_K(x)$ in Fig. \ref{fig-FieldCollision}.  
Since the theory is Lorentz invariant, a kink moving to the left at a speed $v$ is obtained by boosting the stationary solution: $\phi_K[\gamma(x+vt)]$. We set up the initial condition for a collision at $t\rightarrow-\infty$ by linearly superpose two kink solutions
\begin{equation}
\phi(x,t) = \phi_K(x) + \phi_K[\gamma(x+vt)]~.
\label{eq-initial}
\end{equation}
The outcome of the collision can be written as
\begin{equation}
\phi(x,t) = \phi_K(x) + \phi_K[\gamma(x+vt)] + h(x,t)~.\label{eq-h}
\end{equation}
where $h(x,t)$ includes all the perturbations generated by the collision. As we will see, $h$ is small because of the suppression of the space-time-area of interaction of the two solitary waves: $\Aint\propto 1/(\gamma v)$ (see Fig.\ref{fig-IntArea} for details). For ultra-relativistic collisions $1/(\gamma v)\ll1$. 

After the fast-moving solitary wave has moved away from the stationary one, we are essentially left with the stationary solitary wave plus perturbations generated by the collision. This spectrum of perturbations includes the shift in the position $\Delta x$ of the stationary solitary wave
\Beq
\phi_K[x+\Delta x(t)]=\phi_K(x)+\Delta x(t)\phi_K'(x)+\hdots
\Eeq
To extract the phase shift from $h(x,t)$, we expand it as
\Beq
h(x,t)=\phi_K'(x)\Delta x(t)+\hdots
\Eeq
where ``$\hdots$'' are terms orthogonal to $\phi_K'$. The phase shift $\Delta x(t)$ can then  be projected from $h(x,t)$ using
\begin{equation}
\Delta x(t) = \frac{\int dx\, h(x,t)\phi_K'(x)}{\int dx\,\phi_K'^2(x)}~.
\label{eq-extract}
\end{equation}
To find $h(x,t)$, we linearize the EOM \eqn{eq-EOM} to obtain 
\begin{equation}
\left[\partial_t^2-\partial_x^2+W_0(x)\right]h=-\Delta W(x,t)h-S(x,t)~, \label{eq-master}
\end{equation}
where
\begin{eqnarray}
W_0(x)&\equiv& V''[\phi_K(x)]~,\\
\Delta W(x,t)&\equiv&V''[\phi_K(x)+\phi_K[\gamma(x+vt)]]-V''[\phi_K(x)]~, \nn
S(x,t) &\equiv& V'[\phi_K(x)+\phi_K[\gamma(x+vt)]\\
& &\qquad-V'[\phi_K(x)]-V'[\phi_K[\gamma(x+vt)]]~.
\label{eq-source}
\end{eqnarray}
$W_0(x)$ is the mass term for perturbations around an isolated stationary kink, $\Delta W(x,t)$ is the change in mass due to the incoming kink, whereas $S(x,t)$ is the external source which is active only when the two solitary waves overlap. We will treat the incoming kink as a time dependent perturbation in the background of the stationary kink.

Before the collision $h=0$. During the overlap, as can be seen from \eqn{eq-master}, both $\Delta W h$ and $S$ become nonzero, sourcing the perturbation $h$. However, since $\Delta W$ is multiplied by $h$, its effect is suppressed compared to the effect of $S$. Now, since $S$ is only active within $\Aint\propto1/(\gamma v)$, the perturbation $h$ generated by it must also be suppressed by $1/(\gamma v)$. Thus we expect the leading order result of collision to be $\propto 1/(\gamma v)$. 

To solve \eqn{eq-master}, we expand $h$ as 
\begin{eqnarray}
h(x,t)=\sum_a G_a(t)f_a(x),
\end{eqnarray}where $\{f_a(x)\}$ is an orthonormal basis obtained from the eigenvalue equation
\begin{eqnarray}
\left[-\partial_x^2+W_0(x)\right]f_a(x)=E_af_a(x)~.\label{eq-eigen}
\end{eqnarray}
As can be easily checked, the ground state is the zero energy mode $E_0=0$ given by $f_0(x)=M^{-1/2}\phi_K'(x)$. In the previous section we used this zero mode $\propto \phi_K'(x)$ to project out the phase shift $\Delta x$ from the general perturbation $h$. 

Since we are only interested in the phase shift $\Delta x$, instead of solving $h$ from \eqn{eq-master}, we multiply Eq.~(\ref{eq-master}) by $\phi_K'$ and integrate with respect to $x$ to get an equation for $\Delta x$ as a function of $t$:
\begin{equation}
\partial_t^2(\Delta x) = 
\frac{\int dx S(x,t)\phi_K'(x)}{\int dx \phi_K'^2(x)}~.
\label{eq-eom0}
\end{equation}
In deriving the above equation, we used \eqn{eq-eigen}, the orthonormality of $\{f_a(x)\}$ and ignored the $\Delta W h$ term since we are only interested in the leading order effects. We can similarly project onto other modes to obtain the entire spectrum, but here we will focus on the zero mode. With the initial condition $\Delta x(t\rightarrow -\infty)=0$, Eq.~(\ref{eq-eom0}) can be integrated to get
\begin{equation}
\Delta x (t)= 
\frac{1}{M}\int_{-\infty}^{t}d\tau (t-\tau)\int dx S(x,\tau)\phi_K'(x)~,
\label{eq-PS1}
\end{equation}
where $M=\int dx \phi_K'^2(x)$. The source $S(x,t)$ is only turned-on during $\Aint$ and is exponentially close to zero otherwise. Hence the space-time area of integration is limited to $\Aint$ shown in Fig. \ref{fig-IntArea}. Now, consider the co-ordinate transformation
\begin{equation}
q=x~, \ p=\gamma(x+v\tau)~;
~dx d\tau=(\gamma v)^{-1}dqdp~.
\end{equation}
Using these co-ordinate transformations, as well as the restriction of the integration range to $\Aint$, we have
\begin{eqnarray}
\Delta x(t) = 
\frac{1}{(\gamma v)M}\int_{\Aint} dqdp(t-\tau(q,p))S(q,p)\phi_K'(q)
~,
\end{eqnarray}
where $S(q,p)=V'[\phi_K(q)+\phi_K(p)]-V'[\phi_K(q)]-V'[\phi_K(q)]$ and $\tau(q,p)= -(q/v)+p/(\gamma v)$. We now move to field space from $q$-space using
$\phi_K'(q)=\sqrt{2V[\phi_K(q)]}$, which can be obtained from \eqn{eq-kink}. A similar expression holds for $p$ as well. Since $\phi$ is a monotonic function of $q$ and $p$, inverses exist and hence we can use $dq=d\phi_1/\sqrt{2V(\phi_1)}$ and $dp=d\phi_2/\sqrt{2V(\phi_2)}$. After a few integrations by parts, we get Eq.~(\ref{eq-shift}). Explicit integration shows that the term $\propto t$ in \eqn{eq-PS1} is zero to ${\cal O}(\gamma v)^{-2}$. Hence, we are finally left with a time independent phase shift as asserted in \eqn{eq-shift}. Since an antikink $\phi_{AK}(x)=\phi_K(-x)$, replacing $p\rightarrow-p$ in the incoming kink profile leads to an identical leading order phase shift for a kink-antikink collision.
\begin{figure}
\begin{center}
\includegraphics[width=6cm]{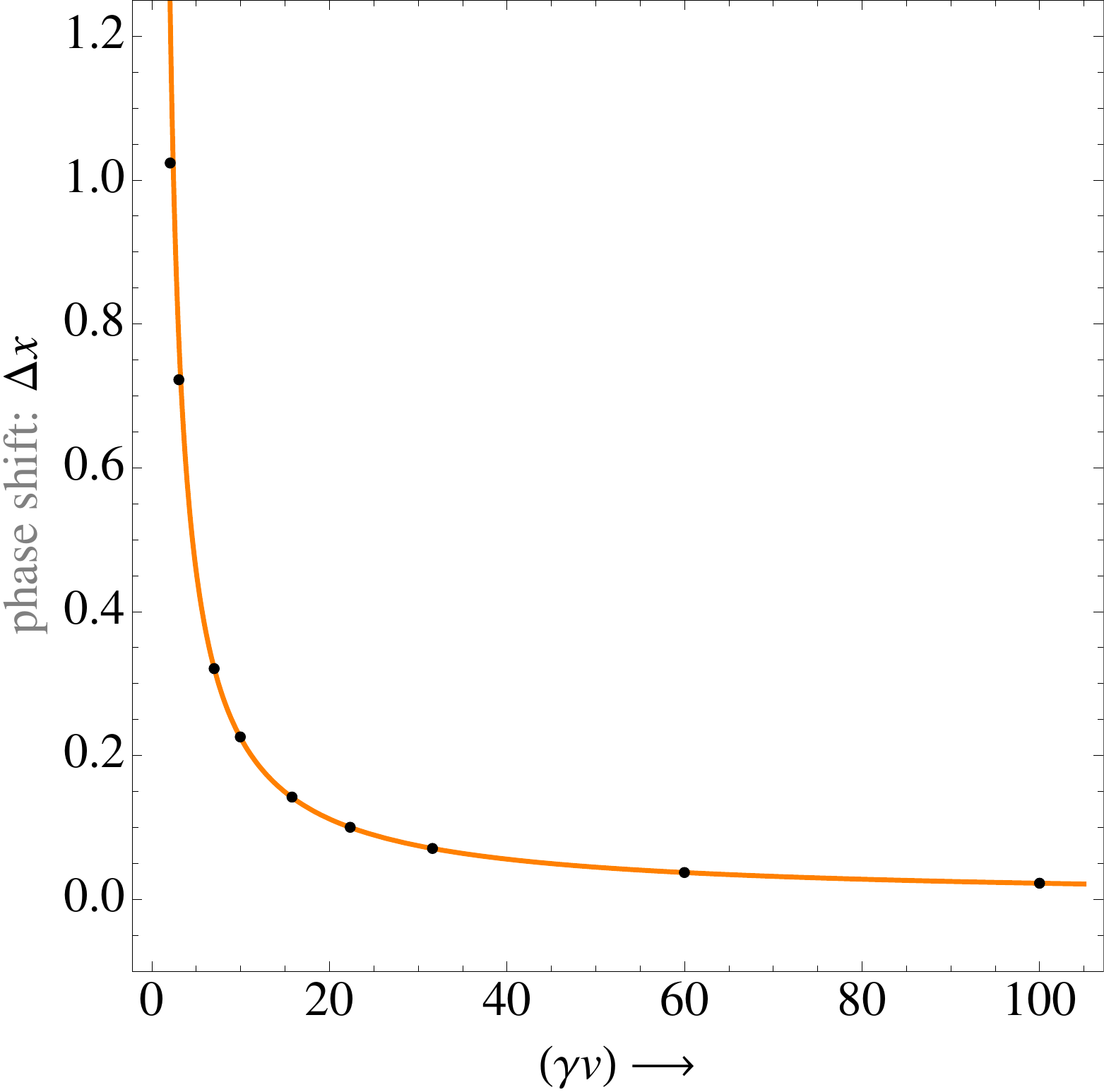}
\caption{We plot the numerically calculated phase shift undergone by a stationary kink colliding with an incoming kink as a function of $(\gamma v)$. For this plot, the scalar field potential $V(\phi)=(1-\cos \phi)(1-0.5\sin^2\phi)$. The orange curve is the theoretical prediction at leading order in $1/(\gamma v)$ and the black dots are the simulation results. 
\label{fig-PhaseShiftG}}
\end{center}
\vskip -20pt
\end{figure}

\section{Comparison with Simulations}
\killspace

To check our expression, we numerically evolve the full equation of motion. From the numerical result $\tilde\phi(x,t)$ we can define
\begin{equation}
\tilde{h}(x,t) = \tilde\phi(x,t) - \phi_K(x)-
\phi_K[\gamma(x+vt)]
\end{equation}
and extract the phase shift from $\tilde{h}$ using Eq.~(\ref{eq-extract}). We use a finite integration range in Eq.~(\ref{eq-extract}) such that the exponential tail of $\phi_K'(x)$ outside is negligible. This range is smaller than the box size where we evolve the full equation of motion, and we make the extraction only after the fast moving kink is sufficiently far from the stationary one.

Note that while the analytical result of $\Delta x$ is time-independent, the numerical result need not be. There are a number of sources of time dependence. First, the orthogonality between $\phi_K'(x)\propto f_0(x)$ and other $\{f_a(x)\}$ is not exact given a finite $x$ integral. Fortunately this additional time dependence is usually periodic and can easily be removed by an average. Another source is a higher order dependence in time---the kink's velocity. It is not surprising that the kink changes velocity after a collision. Our analytical calculation shows that the leading order velocity change is zero, but the next order is generally not. Our numerical data indeed shows that as $(\gamma v)$ increases, the linear dependence drops faster than the constant piece. We fit the slope of this linear dependence and remove it from the constant piece. It is this constant piece that is compared with Eq.~(\ref{eq-shift}).

We applied both the analytical and the numerical methods to the collision of kinks in the model defined by 
\Beq
V(\phi)=(1-\cos \phi)(1-\alpha \sin^2\phi)~,~ -1<\alpha<1~.
\label{eq-pot}
\Eeq 
We carried out a large number of detailed numerical simulations with $-1<\alpha<1$ and $3\leq(\gamma v)\leq100$. The parameter range we cover is clearly beyond small deformation from the Sine-Gordon model ($\alpha=0$) \cite{Mal85b}. While we have chosen a potential that is symmetric around each minimum in the detailed numerical example, we hasten to add that Eq.~(\ref{eq-shift}) does not rely on this symmetry nor this particular form.

\begin{figure}
\begin{center}
\includegraphics[width=6cm]{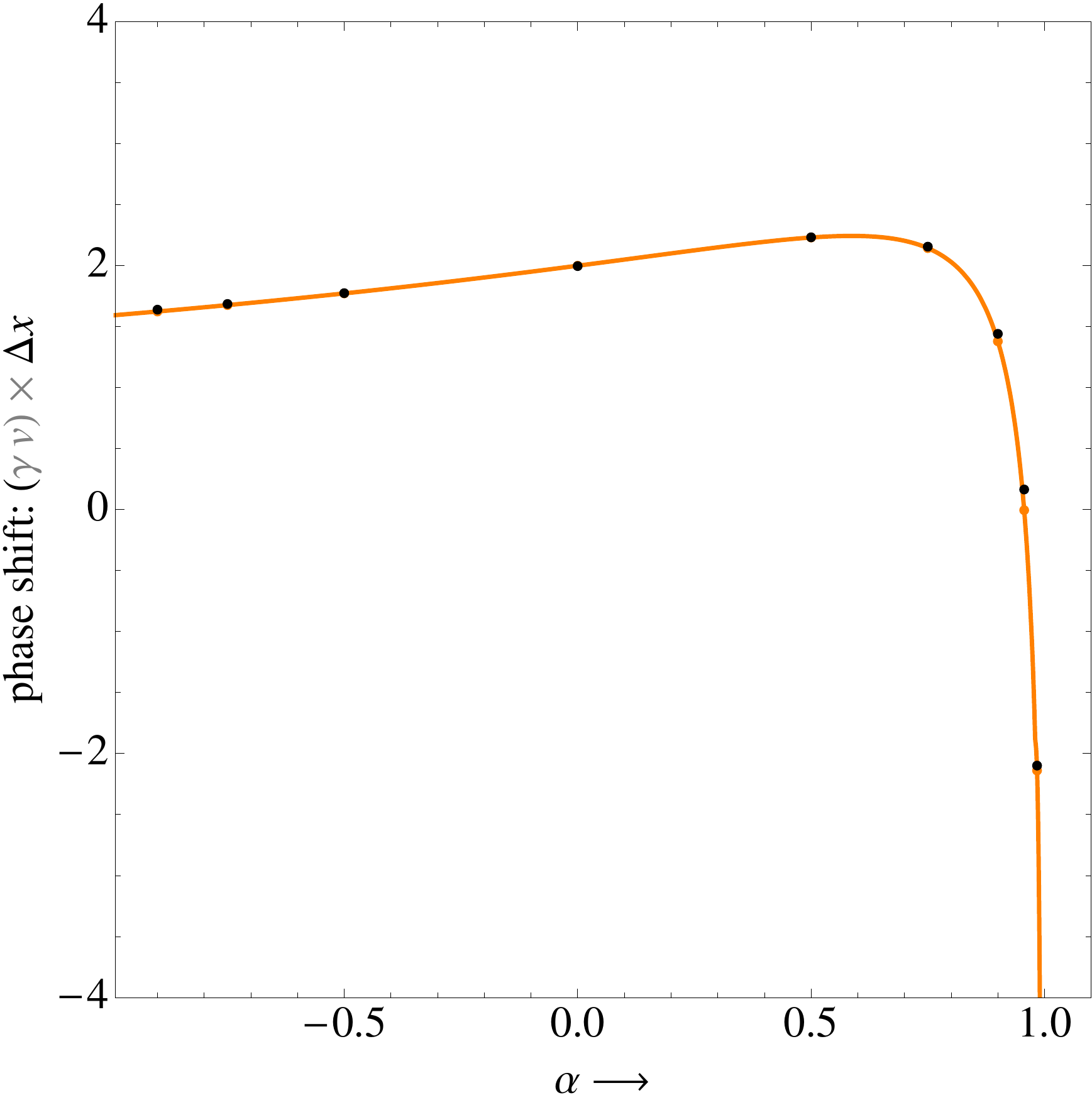}
\caption{The phase shift (multiplied by $(\gamma v)$) undergone by a stationary kink colliding with an incoming kink as a function of the $\alpha$ parameter in the potential $V(\phi)=(1-\cos \phi)(1-\alpha\sin^2\phi)$. For this plot $(\gamma v)=100$. The orange curve (and orange dots) is the theoretical prediction at leading order in $(\gamma v)^{-1}$ and the black dots are the simulation results. 
}\label{fig-PhaseShiftE}
\end{center}
\vskip -20pt
\end{figure}

In Fig. \ref{fig-PhaseShiftG} we plot $\Delta x$ as a function of $(\gamma v)$, and in Fig. \ref{fig-PhaseShiftE} as a function of $\alpha$. Both figures demonstrate the excellent agreement between analytical predictions and numerical results. Note that when $\alpha\gtrsim 0.9564$, the phase shift becomes negative.

\section{Conclusion}
\killspace

Understanding soliton interactions has been an active area of research for more than 50 years. Many interesting physical phenomena involve solitons such as fluxons in Josephson junctions \cite{McLSco78}, non-linear optical solitons \cite{BulCau80}, reheating after inflation \cite{Amin:2011hj} and domain wall collisions in cosmology \cite{EasGib09}. There are two standard approaches. One is to model them as being perturbatively close to the integrable Sine-Gordon system. Another approach is to employ direct numerical simulations. Here, and in our companion paper \cite{ALYlong}, we demonstrate a novel third method -- a kinematics based scattering theory at relativistic velocities. This approach is complementary to those two standard techniques. Our method works well for collisions at ultra-relativistic velocities, which is exactly when numerical descriptions become inefficient. For these collisions, we do not rely on a small deformation from Sine-Gordon, thus our analytical framework is applicable to a wider range of phenomena.

{\it Acknowledgements.} 
We thank Adam Brown, Cliff Burgess, Tom Giblin, Nick Manton, Theodorus Nieuwenhuizen, and especially Erick Weinberg for useful comments. EAL acknowledges the support of an FQXi minigrant. MA thanks the organizers of the Primordial Cosmology workshop at KITP, Santa Barbara, 2013. Numerical simulations were performed on the COSMOS supercomputer, part of the DiRAC HPC, a facility which is funded by STFC and BIS.

\bibliography{all}

\end{document}